\documentclass[prd,preprint,eqsecnum,nofootinbib,amsmath,amssymb,
               tightenlines,dvips]{revtex4-1}
 
\usepackage{url}
\usepackage[dvips]{graphicx}

\newcommand{\gton}{\mathrel{\lower.9ex \hbox{$\stackrel{\displaystyle 
>}{\sim}$}}} 
\newcommand{\lton}{\mathrel{\lower.9ex \hbox{$\stackrel{\displaystyle 
<}{\sim}$}}}  

\begin{document}
                                                                              
\vspace*{1cm}
                                                                              
\title{Hadron Resonance Gas with Repulsive Interactions and  Fluctuations of Conserved Charges}

\author{Pasi Huovinen}
\affiliation
    { Institute of Theoretical Physics, University of Wroclaw, 50204 Wroc\l aw, Poland
    }%
\author{Peter Petreczky}
\affiliation
    { Physics Department,  Brookhaven National Laboratory, Upton, NY 11973, USA
    }%
\date{\today}
                                                                              
\begin{abstract}
We discuss the role of repulsive baryon-baryon interactions in a
hadron gas using relativistic virial expansion and repulsive mean field
approaches. The fluctuations of the baryon number as well as
strangeness-baryon correlations are calculated in the hadron resonance gas
with repulsive interactions and compared with the recent lattice QCD
results. In particular, we calculate the difference between the second
and fourth order fluctuations and correlations of baryon number and
strangeness, that have been proposed as probes of deconfinement. We
show that for not too high temperatures these differences could be
understood in terms of repulsive interactions.
\end{abstract}

\maketitle

\section {Introduction}
Fluctuations and correlations of conserved charges, e.g. baryon number
($B$), electric charge ($Q$) and strangeness ($S$) have been studied
in lattice QCD for some time now. The reason is that they are
sensitive probes of deconfinement and can also be used to calculate
thermodynamic quantities at non-zero baryon density via Taylor
expansion (see Refs.~\cite{Petreczky:2012rq,Ding:2015ona} for recent reviews and
references therein). At sufficiently low temperatures QCD
thermodynamics is expected to be fairly well described by a gas of
non-interacting hadrons and hadron resonances, by so-called hadron
resonance gas (HRG) model~\cite{Hagedorn:1965st}. This picture
naturally emerges from the S-matrix based relativistic virial
expansion, where the interactions are manifested as the phase shifts
of two particle
scattering~\cite{Dashen:1969ep,Venugopalan:1992hy,Weinhold:1997ig,Lo:2017sde}.
In pion-pion and pion-nucleon interactions the repulsive part
associated with the negative phase shifts, is largely canceled by
parts of the positive phase shifts associated with attractive
interactions. The effect of the remaining attractive interactions on
thermodynamics, can be well approximated as a contribution of free
resonances with zero widths~\cite{Venugopalan:1992hy}, although some
differential observables may require explicit treatment of the
interactions~\cite{Huovinen:2016xxq}.

As the temperature increases, particle densities increase, and the
virial expansion only up to second virial coefficient becomes less and
less reliable. To establish the validity of the HRG model at
temperatures close to the QCD transition temperature requires a
detailed comparison with the results from lattice QCD. Early
comparisons have been discussed in
Refs.~\cite{Karsch:2003vd,Karsch:2003zq,Ejiri:2005wq,Cheng:2008zh,Huovinen:2009yb},
where, however, large cutoff effects and/or unphysical quark masses
made a detailed comparison difficult (see e.g.~Ref.~\cite{Huovinen:2009yb}).
In the past several years the fluctuations and correlations of
conserved charges have been studied on the lattice using stout and
highly improved staggered quark (HISQ) actions, and physical quark
masses~\cite{Borsanyi:2010bp,Borsanyi:2011sw,Bazavov:2012jq,Bazavov:2013dta,Bazavov:2013uja,Bazavov:2014xya,Bazavov:2014yba,Borsanyi:2014ewa,Bellwied:2015lba,Ding:2015fca,Bazavov:2015zja,Mukherjee:2015mxc,DElia:2016jqh,Bazavov:2017dus}.
These lattice formulations significantly reduce the cutoff effects. As
the result the comparison between the lattice results and HRG have
become straightforward. Second order fluctuations and correlations
seem to agree reasonably well with the HRG model. However, higher
order fluctuations show deviations from the HRG model close to the
transition temperature. In Ref.~\cite{Bazavov:2013uja} it was argued
that the apparent breakdown of HRG when describing certain differences
of fourth and second order fluctuations and correlations is a signal
of deconfinement. On the other hand, it has been recently shown that the repulsive
interactions modeled by excluded volume can have significant effect on
thermodynamic observables, in particular on higher order
fluctuations~\cite{Albright:2015uua,Vovchenko:2016rkn,Vovchenko:2017zpj}.
The role of repulsive interaction in the context of statistical hadronization
has also been discussed, see e.g.  Ref. \cite{PBM99}.

The aim of this paper is to study the effect of repulsive
baryon-baryon interactions using the S-matrix based virial expansion
and the repulsive mean field approach. In this paper we will
calculate the fluctuations and correlations of conserved charges
defined as
\begin{eqnarray}
&
\displaystyle
\chi_n^X=\left . T^n \frac{\partial^n (p(T,\mu_X)/T^4)}{\partial \mu_X^n} \right |_{\mu_X=0},\\[2mm]
&
\displaystyle
\chi_{nm}^{XY}=\left . T^{n+m} \frac{\partial^{n+m} (p(T,\mu_X,\mu_Y)/T^4)}{\partial \mu_X^n \partial \mu_Y^m}\right |_{\mu_X=0,\mu_Y=0}.
\end{eqnarray}
Here $X=B,Q,S$, i.e.~we consider fluctuations and correlations of
conserved charges corresponding to baryon number, electric charge and
strangeness. It may not be easy to disentangle the effects of repulsive
interactions from other medium effects such as in-medium mass shift
and broadening of width. Therefore it is useful to study the
differences of fluctuations and correlations, which are not affected
by the latter effects. In particular we evaluate $\chi_2^B-\chi_4^B$,
and $\chi_2^B-\chi_6^B$, and show that the inclusion of the repulsive
baryon baryon interaction can naturally explain the temperature
dependence of these differences.

\section{Repulsive interaction in nucleon gas}

First we would like to study the role of repulsive interactions in the
gas of nucleons at temperature $T=1/\beta$. The most natural way to
do this is to consider the virial expansion. In this case the nucleon
pressure can be written as
\begin{equation}
p(T,\mu)=p_0(T)\cosh(\beta \mu)+2 b_2(T) T \cosh(2 \beta \mu).
\end{equation}
Here 
\begin{equation}
p_0(T)=\frac{4 M^2 T^2}{\pi^2} K_2(\beta M)
\end{equation}
is the pressure of free nucleon gas at zero chemical potential and the second
virial coefficient can be written as 
\begin{equation}
b_2(T)=\frac{2 T}{\pi^3} \int_0^{\infty} dE (\frac{ME}{2}+M^2) K_2\left(2 \beta \sqrt{\frac{M E}{2}+M^2}\right)\frac{1}{4 i}
{\rm Tr} \left[ S^{\dagger}\frac{d S}{d E}-\frac{d S^{\dagger}}{dE} S \right],
\end{equation}
with $S$ being the scattering S-matrix and $E$ is the kinetic energy
in the lab frame. Furthermore, $M$ is the nucleon mass and $K_2(x)$
is the Bessel function of second kind. The nucleon-nucleon ($NN$)
interactions break the simple factorisation of the pressure into
temperature dependent and $\mu$-dependent parts. As the result
$\chi_2^B-\chi_4^B$ is not zero as in the case of non-interacting HRG.
Even if the effect of $NN$ interactions is small for the pressure when
$\mu=0$, it could be significant for higher order fluctuations as each
derivative in $\mu$ will multiply $b_2$ by factor two. Because of the
exponential suppression of $K_2(x)$ at large values of the argument,
the virial coefficient $b_2$ is very small for the nucleon
gas. Therefore, it makes sense to introduce the reduced virial
coefficient
\begin{equation}
\bar b_2(T)=\frac{2 T b_2(T)}{p_0(T) K_2(\beta M)}.
\end{equation}
The pressure can now be written as 
\begin{equation}
p(T,\mu)=p_0(T) ( \cosh(\beta \mu)+\bar b_2(T) K_2(\beta M) \cosh(2 \beta \mu) ).
\label{red_vir}
\end{equation}

To evaluate $b_2(T)$ we need to know the S-matrix for the $NN$
scattering. Through the partial wave analysis we have a good
parametrisation of the elastic part of the S-matrix, however, the
inelastic part of the S-matrix is not known. The inelastic channels
open up for $E>280$ MeV and become significant for $E>400$ MeV, and
their importance increases with the energy. We estimate
$b_2(T)$ using the elastic part of the S-matrix and try to include the
effects of the inelastic channel as a systematic uncertainty.

The elastic S-matrix is block diagonal with matrix elements $S_J$,
that are $2\times 2$ matrices for each value of angular momentum
$J$. In the so-called BASQUE parametrisation~\cite{Bugg} $S_J$ has
diagonal elements
\begin{equation}
S_{\pm}=\cos^2 \rho_{\pm}^J \cos 2 \epsilon^J \exp(2 i \delta_{\pm}^J)
\end{equation}
corresponding to orbital angular momenta $L=J\pm 1$, and off-diagonal
elements
\begin{equation}
S_0=i \cos \rho_{+}^J \cos \rho_{-}^J \sin 2 \epsilon^J
 \exp(i(\delta_{+}^J+\delta_{-}^J+\phi^J).
\end{equation}
Here $\delta_{\pm}^J$ are the phase shifts corresponding to angular momentum $J$.
The parameters $\rho_{\pm}^J$ describe the in-elasticity of the
collisions, while $\epsilon^J$ and $\phi^{J}$ are the elastic and
inelastic mixing parameters of $L=J\pm 1$ states.  For $E<280$ MeV the
parameters $\rho_{\pm}^J$ and $\phi^J$ are zero. In this case
\begin{equation}
\frac{1}{4 i}
{\rm Tr} \left[ S^{\dagger}\frac{d S}{d E}-\frac{d S^{\dagger}}{dE} S \right]=
\sum_{s=\pm} \sum_J (2J+1) \left( \frac{d \delta^{J,I=0}_s}{dE} + 3 \frac{d \delta^{J,I=1}_s}{dE} \right),
\label{elonly}
\end{equation}
where we distinguish the isospin zero ($I=0$) and isospin one ($I=1$)
channels in the nucleon-nucleon system. If the parameters
$\rho_{\pm}^J$ and $\phi^J$ are different from zero, the above
equation will become complex, leading to complex value of $b_2(T)$,
which is clearly unphysical. The reason for this problem is that $S_J$
is not unitary. If the inelastic channels were included the unitarity
would be restored, the imaginary terms in the above equation would
drop out and the derivative of inelastic phase shift would
appear. This is easy to see for the simple case when the S-matrix has
one elastic and one inelastic channel~\cite{Lo:2017sde}. In the
following we will set the parameters $\rho_{\pm}^J$ and $\phi$ to zero
and use Eq.~(\ref{elonly}) for all energies to evaluate $b_2$.

In our numerical analysis we use the elastic phase shifts from the
SM16 partial wave analysis~\cite{Workman:2016ysf}. We also use SP07
partial wave analysis~\cite{Arndt:2007qn} as well as an old analysis
from Ref.~\cite{Arndt:1986jb}. The differences arising from the use of
different partial wave analyses are small compared to other
uncertainties of our calculations. For $E>10$ MeV the effects of
Coulomb interactions are small, so the $I=1$ phase shifts are taken
from $pp$ scattering data, while the $I=0$ phase shifts are taken from
the $np$ scattering data. At lower energies the electromagnetic
effects are important and there is a difference between $pp$ phase
shifts and $I=1$ $np$ phase shifts. Since in our study we do not include
electromagnetic interactions for $E<10$ MeV we use the phase shifts
from $np$ scattering data for both $I=0$ and $I=1$ channels. Here it
is sufficient to consider the lowest partial waves ($^1S_0$ for $I=1$
and $^3S_1$ for $I=0$). Finally to obtain the correct threshold
behaviour we use effective range expansion for the S-wave $np$ phase
shifts: $\cot \delta^I=-1/(a_I k)+r_I^0 k/2$, with $a_{I=1}=-23.7$ fm
and $r_{I=1}^0=2.76$ fm for $I=1$~\cite{Arceo}, and $a_{I=0}=5.4194$
fm and $r_{I=0}^0=1.7536$ fm~\cite{deSwart:1995ui}. We checked that
the effective range expansion with the above parameters matches
smoothly to SM16 analysis for $E$ of about few MeV. We note that there
is a large cancellation between the contributions of $I=0$ and $I=1$
channels to $b_2$ at low energies. This is due to different sign of
the scattering length $a_I$ in these two channels and unnaturally
large value of $a_{I=0}$. At high energies the derivative of the sum
of all the phase shifts is negative, which is reflective the repulsive
hard core $NN$ interactions.

\begin{figure}
\includegraphics[width=9cm]{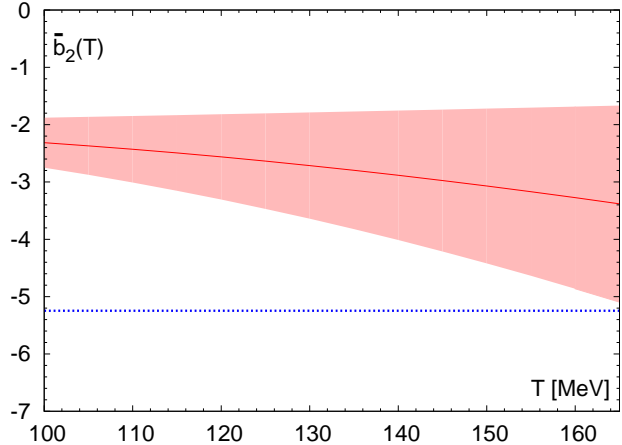}
\caption{The reduced virial coefficient as function of the temperature (solid line) together
with its uncertainty show as the red band (see text). The dashed line corresponds to 
$K M^2/\pi^2$ with $K=450$ MeV fm$^3$.}
\label{fig:b2}
\end{figure}

Finally we need to estimate the uncertainty in $b_2$ due to the
inelastic channels. For this we consider the ratio of the inelastic
to total $pp$ cross-section from SM16 partial wave analysis. The
inelastic cross-section is very small for $E<400$ MeV. For
$400~{\rm MeV} < E < 500~{\rm MeV}$ the inelastic cross-section is about
$10\%$ of the total cross-section. For
$500~{\rm MeV} < E < 600~{\rm MeV}$, $600~{\rm MeV} < E < 800~{\rm MeV}$
and $E > 800~{\rm MeV}$ the inelastic cross-section is about $25\%$,
$40\%$ and $50\%$ of the total cross-section, respectively. Therefore,
we estimate that the uncertainty in $b_2$ that comes from the energy
range $400-500$ MeV, $500-600$, $600-800$ MeV and $>800$ MeV is
$20\%$, $50\%$, $80\%$ and $100\%$, respectively. Here we tried to be
conservative and assumed that the contribution of the (unknown)
inelastic phase shifts to $b_2$ is by a factor two larger than to the
total cross-section. Our numerical result for the reduced virial
coefficient and its uncertainty is shown in Fig.~\ref{fig:b2}.

So far we only considered nucleon-nucleon interactions. Nucleons
can also interact with anti-nucleons. Much less is known about
the interactions between the nucleons and anti-nucleons, but one may
expect that these interactions are significant as well. Fortunately,
the nucleon anti-nucleon interactions give a contribution to the pressure,
which is independent of the chemical potential. Therefore, these interactions
will not affect the fluctuations and correlations that is the main focus of
this paper.

Another way to include the repulsive interaction is via a repulsive
mean field. In this approach it is assumed that the repulsive
interactions lead to shifts in the single particle energies by
$U=K n_b$ and $\bar U=K \bar n_b$ for nucleons and anti-nucleons,
respectively~\cite{Olive:1980dy,Olive:1982we}. Here $n_b$ and
$\bar n_b$ are the densities of nucleons and anti-nucleons defined as
\begin{equation}
n_b=4 \int \frac{d^3 p}{(2 \pi)^3} e^{-\beta(E_p-\mu+U)},~~
\bar n_b=4 \int \frac{d^3 p}{(2 \pi)^3} e^{-\beta(E_p+\mu+\bar U)},~E_p^2=p^2+M^2,
 \label{densities}
\end{equation}
with $\mu$ being the chemical potential corresponding to the net nucleon
density. We use Boltzmann approximation because the nucleon mass is
much larger than the temperature. The phenomenological parameter $K$
characterises the strength of the repulsive interactions and can be
related to the integral of the $NN$ potential over the spatial
volume~\cite{Olive:1980dy,Olive:1982we}. The presence of the short
distance repulsive core in the $NN$ potential implies that $K>0$. 
Requiring, that $\partial p/\partial \mu$ should give the net nucleon
density, i.e $n_b-\bar n_b$ one obtains the following expression for the pressure
\cite{Olive:1980dy,Olive:1982we}
\begin{equation}
p(T,\mu)=T(n_b+\bar n_b)+\frac{K}{2} (n_b^2+\bar n_b^2).
\end{equation}
In principle Eq.~(\ref{densities}) should be solved self-consistently
to obtain $n_b$ ($\bar n_b$). However, for temperatures below the QCD
transition temperature $n_b$ ($\bar n_b$) is small, and for typical
phenomenological values of $K$, e.g.
$K=450$ MeV fm$^3$~\cite{Sollfrank:1996hd}, $\beta U$ is small
too. For example even for $T=175$ MeV we get $\beta U=0.077$.
Therefore we can expand the exponential in the equations for $n_b$ and
$\bar n_b$, and the factor $(1+n_b)^{-1}$ and $(1+\bar n_b)^{-1}$ when solving $n_b$ and $\bar n_b$,
and write
\begin{equation}
n_b=n_b^0 (1 -\beta K n_b^0), ~\bar n_b=\bar n_b^0 (1 -\beta K \bar n_b^0),
\end{equation}
with $n_b^0$ and $\bar n_b^0$ being the free nucleon and anti-nucleon
densities. With this the pressure can be written in terms
of $n_b^0$ and $\bar n_b^0$ as follows:
\begin{equation}
p(T,\mu)=
T(n_b^0+\bar n_b^0)-\frac{K}{2} \left(\left(n_b^0\right)^2
+\left(\bar n_b^0\right)^2 \right).
\end{equation}
Taking into account that $n_b^0=2 T M^2/\pi^2 K_2(\beta M) e^{\beta\mu}$
and $\bar n_b^0=2 T M^2/\pi^2 K_2(\beta M) e^{-\beta\mu}$
we finally get
\begin{equation}
p(T,\mu)=\frac{4 T^2 M^2}{\pi^2} K_2(\beta M) \cosh(\beta \mu)-4 K \frac{T^2 M^4}{\pi^4} K_2^2(\beta M)\cosh(2 \beta \mu)
\end{equation}
The structure of the above equation is very similar to the one
obtained in the virial expansion. The correction to the free gas
result is negative and the factorisation of the pressure in
$T$-dependent part and $\mu$ dependent part does not hold. Comparing
the above result with the virial expansion result one can determine
the value of $K$ at some temperature. To estimate the relative size of
the second term in the above equation we write
\begin{equation}
p(T,\mu)=p_0(T) ( \cosh(\beta \mu)-\frac{K M^2}{\pi^2} K_2(\beta M) \cosh(2 \beta \mu) ).
\end{equation}
Comparing this equation with Eq.~(\ref{red_vir}) we see that $-K M^2/\pi^2$ corresponds
to the reduced virial coefficient $\bar b_2(T)$. Therefore, in Fig.~\ref{fig:b2} we show
this combination for the previously used phenomenological value $K=450$ MeV fm$^3$. At low 
temperatures $-\bar b_2(T)$ is significantly smaller than $K M^2/\pi^2$. However, at the
highest temperatures the two agree. We stress again that the smallness of $-\bar b_2$ comes
from the cancellation of positive and negative contributions in the $I=0$ and $I=1$
channels. Such cancellation is a somewhat accidental feature of the $NN$ interactions
and may not be present for other baryons. For these reasons we will use the value
$K=450$ MeV fm$^3$ in what follows.

Finally, we note that the first quantum correction to the pressure of the nucleon gas is
$-M^2 T^2/\pi^2 K_2(2 \beta M) \cosh(2 \beta \mu)$. It has the same dependence on $\mu$
as the contribution of repulsive interactions but is about $20$ times smaller. Therefore,
it will be neglected in the following analysis.

\section{Repulsive mean field in multi-component hadron gas and fluctuations
of conserved charges}

It is straightforward to generalise the above repulsive mean field approach
to multi-component hadron gas. The baryon density is written as
\begin{equation}
n_B(T,\mu_B,\mu_S,\mu_Q)=\frac{T}{2 \pi^2} \sum_i g_i M_i^2 K_2(\beta M_i) e^{\beta \mu_{i,eff}},
\label{nB}
\end{equation}
where $M_i$ is the mass of the $i^{th}$ baryon and $g_i$ is the
corresponding degeneracy factor. Furthermore, the effective chemical
potential of the $i^{th}$ baryon is given by
\begin{equation}
\mu_{i,eff}=\sum_j q_i^j \mu_j - K n_B,
\end{equation}
with $(q_i^1,q_i^2,q_i^3)=(B_i,S_i,Q_i)$ being the baryon number, strangeness and
electric charge of the $i^{th}$ baryon. Here we assumed that the
repulsive interaction is the same for all baryons. This is clearly an
oversimplification. While lattice calculations indicate that repulsive
core in the central potential is similar for many baryon combinations
(e.g. $NN$, $\Lambda N$, $\Lambda \Lambda$, etc.), there are some differences~\cite{Doi:2017cfx}.
The hyperon nucleon and hyperon-hyperon interactions have been studied
also in chiral effective theory \cite{Polinder:2006zh,Haidenbauer:2013oca}. 
It has been found that these interactions are dominantly repulsive but 
different from nucleon-nucleon interactions.
However, we do not have sufficient information about baryon-baryon
interactions to come up with a more sophisticated mean field model.
Replacing $\mu_{i,eff}$ in Eq.~(\ref{nB}) by
$\bar \mu_{i,eff}= \sum_j \bar q_i^j \mu_j - K \bar n_B$, we obtain the
density of anti-baryons, $\bar n_B$. Note that $\bar q_i^j=-q_i^j$.
Expanding the exponential to leading order
in $K$ as in the previous section for the baryon and antibaryon
densities, and again requiring that $\partial p/\partial \mu_B = n_B - \bar n_B$,
we obtain
\begin{equation}
p_B(T,\mu_B,\mu_S,\mu_Q)=T(n_B^0+\bar n_B^0)-\frac{K}{2} \left(\left(n_B^0\right)^2
+\left(\bar n_B^0\right)^2 \right),
\end{equation}
where $n_B^0$ and $\bar n_B^0$ are the free baryon and anti-baryon densities.
The pressure of the free baryon gas can be decomposed into partial
baryonic pressure of strangeness one, strangeness two, and strangeness
three baryons, and the same is true for anti-baryons. Therefore, we
write
\begin{eqnarray}
&
\displaystyle
p_B(T,\mu_B,\mu_S,\mu_Q)=\tilde p_B(\mu_S,\mu_Q) e^{\beta \mu_B} + \tilde p_B(-\mu_S,-\mu_Q) e^{-\beta \mu_B}\nonumber\\
&
\displaystyle
-\frac{\beta^2 K}{2} \left( \tilde p_B^2\left(\mu_S,\mu_Q\right) e^{2 \beta \mu_B} + \tilde p_B^2\left(-\mu_S,-\mu_Q\right) e^{-2\beta \mu_B} \right),
\end{eqnarray}
with
\begin{equation}
\tilde p_B(\mu_S,\mu_Q)=p_B^{S0}+p_B^{S1} e^{-\beta \mu_S}+p_B^{S2} e^{-2 \beta \mu_S}+p_B^{S3} e^{-3 \beta \mu_S},
\end{equation}
and $p_B^{Sk}$ denotes the contribution of $S=-k$ baryons to the free pressure at zero chemical
potentials. With this it is straightforward to get the baryon number fluctuations and
baryon-strangeness correlations
\begin{eqnarray}
\displaystyle
\chi_n^B&=&\chi_n^{B(0)}-2^n \beta^4 K \left(N_B^0\right)^2, \hspace*{37mm}(n~\mathrm{even})
\label{chiB}\\
\displaystyle
\chi_{n1}^{BS}&=&\chi_n^{BS(0)}+2^{n+1} \beta^5 K N_B^0 (p_B^{S1}+2 p_B^{S2}+3 p_B^{S3}) .
\qquad(n~\mathrm{odd})\label{chiBS}
\end{eqnarray}
Here 
\begin{equation}
N_B^0(T)=\frac{T}{2 \pi^2} \sum_i g_i M_i^2 K_2(\beta M_i)
\end{equation}
and the subscript "0" in the above equation refers to the non-interacting HRG.

In Ref.~\cite{Bazavov:2013dta} it was suggested that certain
combinations of fluctuations and correlation of conserved charges can
be used as indicators of deconfinement. In particular, the following
two combinations
\begin{equation}
\chi_{31}^{BS}-\chi_{11}^{BS},\qquad\mathrm{and}\qquad\chi_2^B-\chi_4^B
\end{equation}
have been suggested as measures of deconfinement in the light and
strange hadron sectors, respectively. In non-interacting HRG these
quantities are identically zero, while they have non-zero values for
the ideal quark gas. The lattice results show that these quantities
quickly rise above zero around the transition temperature and start
approaching the ideal gas limit for $T>200$ MeV. This was interpreted
as a transition from non-interacting hadron gas to quark
gas~\cite{Bazavov:2013dta}. Therefore, it is interesting to see to
what extent the increase in $\chi_{31}^{BS}-\chi_{11}^{BS},\qquad\mathrm{and}\qquad\chi_2^B-\chi_4^B$
around the transition temperature can be explained with the repulsive baryon
interactions.

We calculated $\chi_2^B-\chi_4^B$, $\chi_2^B-\chi_6^B$ and
$\chi_{31}^{BS}-\chi_{11}^{BS}$ in the HRG model with repulsive mean
field using Eqs.~(\ref{chiB}) and (\ref{chiBS}). We considered only
the contribution of ground state octet and decuplet baryons. The
excited baryon states should appear as attractive (resonant)
interactions in the hadron gas and thus, they are included in the
non-interacting part of HRG. On the other hand, when resonances are
interpreted as arising from attractive interactions, they lead to an
increase in the density of ground state
baryons~\cite{Weinhold:1997ig}. We leave creating a proper treatment
of heavy resonances for a further study \cite{new}, and, as mentioned,
concentrate here on the effects of ground state baryons and the lowest
resonances.

\begin{figure}
\includegraphics[width=9cm]{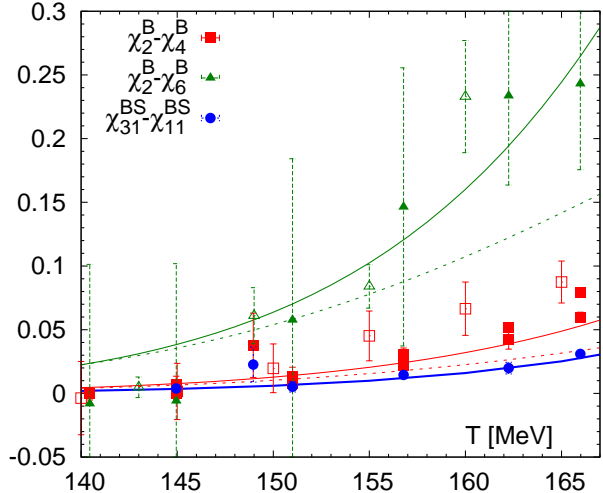}
\caption{The differences $\chi_{31}^{BS}-\chi_{11}^{BS}$, 
  $\chi_2^B-\chi_4^B$ and $\chi_2^B-\chi_6^B$
  calculated in the HRG model with
  repulsive mean field (dotted, solid and dashed lines) and in lattice QCD. The filled symbols
  correspond to lattice calculations of $\chi_2^B-\chi_4^B$ and
  $\chi_2^B-\chi_6^B$ with HISQ action on $32^3 \times 8$
  lattices~\cite{Bazavov:2017dus}. The open symbols correspond to lattice
  results on $\chi_2^B-\chi_4^B$~\cite{Bellwied:2015lba} as well as to
  lattice results on $\chi_2^B-\chi_6^B$~\cite{DElia:2016jqh}. For
  $\chi_{31}^{BS}-\chi_{11}^{BS}$ the lattice results from
  Ref.~\cite{Bazavov:2013dta} are used. The dashed lines correspond to the 
  unexpanded mean field result (see text).}
\label{fig:comp2lat}.
\end{figure}

As discussed before we use the value $K=450$ MeV fm$^3$ in our
numerical study. Our results are shown in Fig.~\ref{fig:comp2lat} and
compared with the lattice results obtained with HISQ
action~\cite{Bazavov:2013dta,Bazavov:2017dus} depicted with filled symbols.
We also use the lattice results for $\chi_2^B-\chi_6^B$ obtained with
stout action~\cite{DElia:2016jqh} as well as continuum extrapolated
results for $\chi_2^B-\chi_4^B$ from
Ref.~\cite{Bellwied:2015lba}, depicted with open symbols. 
As expected the effect of the repulsive interactions is bigger for
$\chi_6^B$ than for $\chi_4^B$. 
In our analysis so far we assumed that the density of baryons(anti-baryons) is
small and therefore we kept the leading order term of the expansion 
in baryon density, i.e. the term proportional to $K$ (c.f. Eqs. (\ref{chiB})
and (\ref{chiBS}) ). As the temperature is increasing the number density of
baryons and anti-baryons also increases and this expansion become less reliable.
Therefore, we also calculated $\chi_2^B-\chi_4^B$ and $\chi_2^B-\chi_6^B$ using
the unexpanded mean-field expressions and the results are shown in Fig. ~\ref{fig:comp2lat}
as dashed lines. The difference between the expanded and un-expanded mean field
results is significant at and above the crossover temperature. The full mean field
result is below the lattice data. This problem could be cured by taking into account
the effect of repulsive interactions for higher baryon resonances, although
it is not clear to what extent the HRG model is reliable in this temperature region.
Note, that using the full mean field result is more important for the higher order
fluctuations and correlations than for the pressure 
since the effect of the repulsive interactions is enhanced by factor $2^n$ for
the former (c.f. Eqs. (\ref{chiB}) and (\ref{chiBS}) ).
In Ref. \cite{Vovchenko:2016rkn} the decrease of
  $\chi_4^B/\chi_2^B$ from one was described in terms of HRG, where
  the repulsive interactions are modeled by excluded volume and good
  agreement with the lattice data was found. The increase in
  $\chi_2^B-\chi_4^B$ is equivalent to decrease of $\chi_4^B/\chi_2^B$
  from unity, and thus our analysis confirms this result.

In Ref.~\cite{Bazavov:2013dta} another combination of strangeness
fluctuations and baryon-strangeness correlation has been considered,
which is identically zero in the ideal HRG and approaches the free
quark gas value at very high temperature, namely
\begin{equation}
v_2=\frac{1}{3} (\chi_2^S-\chi_4^S)-2 \chi_{13}^{BS}-4 \chi_{22}^{BS}-2 \chi_{31}^{BS}.
\end{equation}
We calculated $v_2$ in our HRG model with repulsive mean field. We
find that it has different sign depending on the value of $K$ and the
temperature, while lattice calculation shows that $v_2$ is positive
and monotonically increases with the temperature. So the simplest mean
field approach with the same mean-field for all baryons cannot
describe this quantity, and the differences in the repulsive baryon
interactions in strange and non-strange baryons are important
here. This is contrary to the difference
$\chi_{31}^{BS}-\chi_{11}^{BS}$ where the repulsive interactions in
the different strangeness sectors contribute with the same sign. To
understand $v_2$ in the framework of the hadron gas with repulsive
interactions more information on baryon-baryon interactions in
different strangeness sectors will be needed.

We also calculated the baryon electric charge correlations $\chi_{31}^{BQ}$
and $\chi_{11}^{BQ}$ in the repulsive mean field approach. The results
are similar to the case of $\chi_{31}^{BS}$ and $\chi_{11}^{BS}$. 
In particular, $\chi_{31}^{BQ}-\chi_{11}^{BQ}$ increases with increasing
temperature and the repulsive interactions between different baryons
contribute with the same sign. Our results agree with the preliminary lattice
results.

\section{Conclusions}

In this paper we discussed the role of repulsive baryon interactions
on the thermodynamics and fluctuations of conserved charges of hadronic
matter using relativistic virial expansion and repulsive mean field
approach. We showed that the two approaches lead to almost identical
results. In particular the reduced virial coefficient $\bar b_2(T)$
shows only a mild temperature dependence and corresponds to the
combination $K M^2/\pi^2$ appearing in the repulsive mean field
approach. The deviations from ideal HRG for higher order fluctuations
and correlations of conserved charges can be naturally explained by
the repulsive interactions. 
We pointed out that it is useful to study the effect
of repulsive interactions in terms of the following 
differences: $\chi_{31}^{BS}-\chi_{11}^{BS}$,
$\chi_2^B-\chi_4^B$ and $\chi_2^B-\chi_6^B$ since the ideal
hadron resonance gas part drops out and thus the results are not sensitive
to the hadron spectrum. This makes it easy to disentangle the effects
of repulsive interactions from other effects such as missing states \cite{Bazavov:2014xya}
and in-medium modifications of hadron properties.
The size of the deviations from
the ideal gas limit  
for these differences obatined
in the simple mean field model is similar to that observed on the lattice,
though the former has large uncertanties at and above the QCD crossover
temperature.
However, not all strangeness baryon correlations can be understood within
our simple mean field approach due to the fact that baryon-baryon
interactions are different in different strangeness sectors.
Therefore, in the future it will be
important to refine the treatment of the repulsive interactions of strange baryons 
using information from lattice QCD 
and chiral effective theory \cite{Polinder:2006zh,Haidenbauer:2013oca}
to obtain a better
description of the fluctuations and correlation of conserved charges.
Nevertheless, it is clear that HRG with repulsive
interactions is a useful approach for studying the contribution of
baryons to the thermodynamics of hadronic matter at zero and not too high
baryon density. 
It was shown in Ref.~\cite{Vovchenko:2016rkn} that including
  repulsive interactions by excluded volume affects the equation of
  state and fluctuations of conserved charges improve the agreement
  with the lattice data. Along similar lines we plan to study the QCD
  equation of state and fluctuations of conserved charges at zero and
  non-zero baryon density using HRG model with repulsive mean field
  \cite{new} and perform detailed comparisons to the available lattice results. 
We hope that this study will also clarify the range of applicability
of the mean field model.

\section*{Acknowledgment}
This work was supported by U.S.Department of Energy under Contract
No.~DE-SC0012704, and by National Science Center, Poland, under grant
Polonez DEC-2015/19/P/\allowbreak ST2/03333 receiving funding from the European
Union's Horizon 2020 research and innovation programme under the Marie
Sk\l odowska-Curie grant agreement No 665778. We thank I.~Strakovsky
for correspondence on partial wave analysis of nucleon-nucleon
scattering and for sending the results for the BASQUE parametrisation.
We thank Sz.~Bors\'anyi, M.~D'Elia and G.~Gagliardi for sending the
lattice results on baryon number fluctuations. PP would like to thank
J. Haidenbauer for useful discussions on the application of the chiral
effective theory to the baryon-baryon interactions.
PH thanks P.M. Lo for illuminating
discussions on the S-matrix approach.

\bibliography{ref}

\end{document}